\documentclass[aps,prd,preprint,nofootinbib,longbibliography,superscriptaddress,eqsecnum,amsmath,amsfonts,amssymb]{revtex4-2}
\usepackage{graphicx} % Required for inserting images
\usepackage{physics}
\usepackage{xcolor}
\usepackage{hyperref}
\usepackage{tikz}
\usepackage{bbm}

\newcommand\nn{\nonumber}

\allowdisplaybreaks[1]

\begin{document}

\title{Symmetries of non-maximal supergravities with higher-derivative corrections}
\preprint{USTC-ICTS/PCFT-26-21}
\date{\today}

\author{Yi Pang}\email{pangyi1@tju.edu.cn}
\affiliation{Center for Joint Quantum Studies and Department of Physics, School of Science, Tianjin University, Tianjin 300350, China}
\affiliation{Peng Huanwu Center for Fundamental Theory, Hefei, Anhui 230026, China}

\author{Robert J. Saskowski}\email{robert\_saskowski@tju.edu.cn}
\affiliation{Center for Joint Quantum Studies and Department of Physics, School of Science, Tianjin University, Tianjin 300350, China}

\begin{abstract}
    We consider hidden symmetries arising from U-duality in the dimensional reduction of non-maximal higher-derivative supergravities to three dimensions. In particular, we consider the $G_{2(2)}$ symmetry of minimal five-dimensional supergravity and the $O(d+p+1,d+1)$ symmetry of bosonic and heterotic string theory on $T^d$. Using a group theory argument, we show that the higher-derivative corrections explicitly break all hidden symmetry enhancements. As special cases, this also implies that higher-derivative corrections prevent the symmetry enhancement to $SL(3,\mathbb R)$ in pure five-dimensional gravity and $O(4,4)$ in the STU model.
\end{abstract}

\maketitle
\vspace{1cm}

\newpage

\tableofcontents

\section{Introduction}
There is a wealth of hidden symmetries that arise upon dimensional reduction of gravitational theories. This idea originates with Ehlers' observation that four-dimensional gravity reduced on a circle has an enhanced $SL(2,\mathbb R)$ symmetry~\cite{Ehlers:1959aug}, which was later extended to the observation that the reduction of five-dimensional gravity on $T^2$ has a hidden $SL(3,\mathbb R)$ symmetry~\cite{Maison:1979kx,Maison:2000fj}. One of the simplest sources of such hidden symmetries is from dualities of string theories. In particular, any string theory reduced on a $d$-dimensional torus $T^d$ will exhibit an $O(d,d)$ symmetry arising from T-duality~\cite{Maharana:1992my,Sen:1994fa}. Heterotic strings have gauge fields that transform under either $E_8\times E_8$ or $SO(32)$. If the background gauge field commutes with a $U(1)^p$ subgroup of the gauge group, then the heterotic T-duality group is enlarged to $O(d+p,d)$. Moreover, specific choices of theories and dimensions lead to larger symmetry enhancements, due to S-duality. Here, we will be interested in reductions to three dimensions. In this case, bosonic strings on $T^{23}$ receive a symmetry enhancement to the U-duality group $O(24,24)$ and heterotic strings on $T^7$ receive a symmetry enhancement to $O(24,8)$~\cite{Sen:1994wr}. Type II supergravity on $T^7$ receives a similar U-duality enhancement to $O(8,8)$, but the presence of RR fields further enlarges this to $E_{8(8)}$~\cite{Cremmer:1997ct,Cremmer:1998px}. We will refer to these generically as $O(d+p+1,d+1)$ symmetry enhancements. Similarly, the reduction on $T^2$ of minimal five-dimensional supergravity yields a $G_{2(2)}$ hidden symmetry~\cite{Cecotti:1988qn,Mizoguchi:1998wv,Cremmer:1999du,Cremmer:1997ct,Cremmer:1998px} and of the STU model yields an $O(4,4)$ hidden symmetry~\cite{Chong:2004na,Galtsov:2008bmt,Galtsov:2008jjb}.

In all these cases, upon dimensional reduction, the resulting theory takes the form of a coset model, described in terms of a scalar metric $\mathcal M$ with target space $G/H$, where $G$ is the hidden symmetry group and $H$ is a maximal (pseudo)compact\footnote{When reducing on a spacelike internal manifold, this will be a maximal compact subgroup; however, when reducing over a timelike direction, this becomes pseudocompact.} subgroup of $G$. In this context, $G$ is the isometry group of the target space, while $H$ acts as the local isotropy subgroup. On the other hand, a solution with a $U(1)^d$ isometry may be viewed as compactified on a $d$-dimensional torus. As such, hidden symmetries can be used to generate new inequivalent solutions in the parent theory by applying a $G$ transformation to the torus moduli of a solution with abelian isometries. Various authors have leveraged the hidden symmetries of $O(d+p,d)$~\cite{Veneziano:1991ek,Meissner:1991zj,Sen:1991zi,Sen:1991cn,Gasperini:1991qy,Hassan:1991mq,Sen:1992ua,Cvetic:1995sz,Cvetic:1995kv,Cvetic:1996xz}, $G_{2(2)}$~\cite{Bouchareb:2007ax,Tomizawa:2008qr,Compere:2009zh,Suzuki:2024coe,Suzuki:2024phv,Suzuki:2024vzq}, and $O(4,4)$~\cite{Chong:2004na,Galtsov:2008bmt,Galtsov:2008jjb} to generate black hole solutions in this way. However, these aforementioned results are all for two-derivative supergravity.

As general relativity and supergravity are non-renormalizable, they are best thought of as the leading order in an effective field theory (EFT) expansion,
\begin{equation}
    \mathcal L=\mathcal L_{2\partial}+\Lambda_c^{-2}\mathcal L_{4\partial}+\mathcal O(\Lambda_c^{-4}),
\end{equation}
where the two-derivative Lagrangian $\mathcal L_{2\partial}$ consists of general relativity minimally coupled to some choice of matter (Standard Model, hidden sector, \emph{etc}.) and the higher-derivative corrections are suppressed by some large cutoff scale $\Lambda_c$. These higher-derivative corrections then encode both the effects of UV physics and quantum corrections. As such, it is of paramount importance to understand the extension of two-derivative results to higher-derivative orders. Supergravity, being the low-energy description of string theory, selects a distinguished choice of EFT, where the cutoff scale $\Lambda_c^{-2}$ is identified with the string scale $\alpha'$. 

Thus, we are generally interested in the extension of hidden symmetries to higher-derivative orders. In the stringy context, the $O(d+p,d)$ T-duality symmetry is known to persist to all orders in the tree-level $\alpha'$-expansion~\cite{Sen:1991zi,Hohm:2014sxa}, and recent work has extended the use of four-dimensional $O(2,1)$ solution generation to the four-derivative level~\cite{Hu:2025aji,Xia:2025lvn}. While T-duality is realized perturbatively, U-duality is expected to be a symmetry of the fully nonperturbative theory~\cite{Sen:1994wr,Hull:1994ys}. As such, one should start with a higher-derivative action that includes all orders of loop and instanton corrections. For example, the full eight-derivative corrections in type IIB supergravity in ten dimensions take the form~\cite{Green:1997tv,Green:1997di,Green:1997as,Kiritsis:1997em,Green:1997me,Pioline:1998mn,Green:1998by,Green:1999pu,Obers:1999es,Sinha:2002zr,Berkovits:2004px,DHoker:2005vch,DHoker:2005jhf,Matone:2005vm,Green:2005ba,Green:2006gt,Basu:2007ru,Basu:2007ck,Garousi:2013qka}
\begin{equation}
    e^{-1}\mathcal L_{8\partial}^\mathrm{IIB}=\frac{\alpha'^3}{3\cdot 2^{13}}E_{3/2}(\tau,\bar\tau)\qty(t_8t_8+\frac{1}{4}\epsilon_8\epsilon_8)R^4+\cdots,\label{eq:SL2invAction}
\end{equation}
where $\tau$ is the complex IIB modulus and $E_{3/2}$ is the $SL(2,\mathbb Z)$-invariant non-holomorphic Eisenstein series. This action is invariant under the $SL(2,\mathbb Z)$ S-duality group. However, this is difficult to study. Fortunately, heterotic and bosonic strings receive corrections starting at four-derivatives
\begin{equation}
    e^{-1}\mathcal L_{4\partial}=e^{-2\phi}R_{\mu\nu\rho\sigma}R^{\mu\nu\rho\sigma}+\cdots,
\end{equation}
which are all-loop exact~\cite{Ellis:1989fi}. In particular, the four-derivative heterotic action is uniquely determined by supersymmetry, up to an overall constant~\cite{Bergshoeff:1989de}. However, a simple scaling argument quickly shows that an $\mathbb R^+$ scaling contained in the U-duality group $O(d+p+1,d+1)$ is broken~\cite{Lambert:2006he,Bao:2007er}. It was recently shown in Ref.~\cite{Eloy:2022vsq} that, in the $p=0$ case, this breaks $O(d+1,d+1)$ to $O(d,d)\ltimes\mathbb R^{2d}$, where $\mathbb R^{2d}$ corresponds to global shifts of the NS two-form.

An even simpler case to study is that of the $G_{2(2)}$ symmetry enhancement of minimal five-dimensional supergravity. There, it is known that there is a unique four-derivative action~\cite{Liu:2022sew}, which necessarily encompasses all orders of quantum corrections. However, a similar scaling argument applies to this four-derivative action. Nevertheless, it is not known in general whether this completely prevents the symmetry enhancement or merely restricts it to a subgroup of the original hidden symmetry. In principle, preserving a subgroup of the hidden symmetry would still allow one to generate a subset of higher-derivative corrected solutions. In this paper, we use a simple group theory argument to show that the presence of higher-derivative corrections explicitly breaks all of the U-duality symmetry enhancement of $G_{2(2)}$. This includes the $SL(3,\mathbb R)$ hidden symmetry of pure five-dimensional gravity as a special case. This is the main novel result of the paper and the focus. However, as a point of comparison, we then apply this argument to the case of the half-maximal $O(d+p+1,d+1)$ symmetry enhancement to rederive the result of~\cite{Eloy:2022vsq} (without truncating the heterotic gauge fields) that the enhancement of $O(d+p,d)$ to $O(d+p+1,d+1)$ is fully broken. This includes the $O(4,4)$ symmetry enhancement of the STU model as a special case.

The rest of this paper is organized as follows. In Section~\ref{sec:G22}, we review the structure of the group $G_{2(2)}$ and its action on the moduli in minimal five-dimensional supergravity. We then show, via a group-theoretic argument, that the breaking of $\mathbb R^+$ scaling symmetry by the higher-derivative corrections necessarily breaks $G_{2(2)}$ to its geometric subgroup. In Section~\ref{sec:Odpd}, we apply this group theoretical argument to the $O(d+p+1,d+1)$ U-duality of the bosonic/heterotic string and show that it is broken to the T-duality subgroup. We conclude with some discussion and future directions in Section~\ref{sec:disc}.

\section{$G_{2(2)}$ U-duality}\label{sec:G22}
\subsection{Group structure of $G_{2(2)}$}
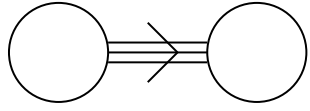
\begin{figure}
    \centering
\tikzset{every picture/.style={line width=0.75pt}} %set default line width to 0.75pt        

\begin{tikzpicture}[x=0.75pt,y=0.75pt,yscale=-1,xscale=1]

%Shape: Circle [id:dp7673705747672235] 
\draw   (110,135) .. controls (110,121.19) and (121.19,110) .. (135,110) .. controls (148.81,110) and (160,121.19) .. (160,135) .. controls (160,148.81) and (148.81,160) .. (135,160) .. controls (121.19,160) and (110,148.81) .. (110,135) -- cycle ;
%Shape: Circle [id:dp19575501074190005] 
\draw   (210,135) .. controls (210,121.19) and (221.19,110) .. (235,110) .. controls (248.81,110) and (260,121.19) .. (260,135) .. controls (260,148.81) and (248.81,160) .. (235,160) .. controls (221.19,160) and (210,148.81) .. (210,135) -- cycle ;
%Straight Lines [id:da8557647511074813] 
\draw    (160,135) -- (210,135) ;
%Straight Lines [id:da7412876111625133] 
\draw    (160,140) -- (210,140) ;
%Straight Lines [id:da8187709803453581] 
\draw    (160,130) -- (210,130) ;
%Straight Lines [id:da8978254358335324] 
\draw    (180,120) -- (195,135) ;
%Straight Lines [id:da6735136612314301] 
\draw    (195,135) -- (180,150) ;
\end{tikzpicture}
\caption{Dynkin diagram of $\mathfrak g_{2(2)}$.}
\label{fig:dynkin}
\end{figure}

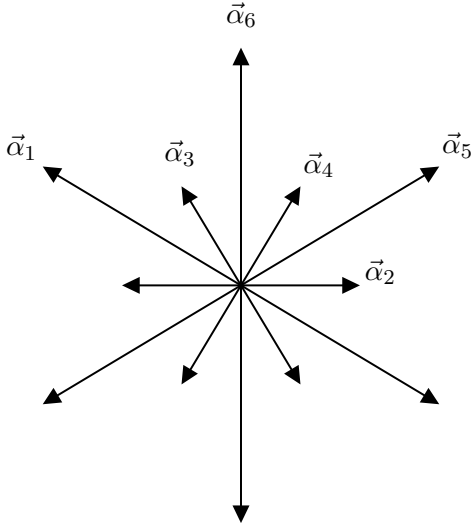
\begin{figure}

\tikzset{every picture/.style={line width=0.75pt}} %set default line width to 0.75pt        

\begin{tikzpicture}[x=0.75pt,y=0.75pt,yscale=-1,xscale=1]

%Straight Lines [id:da3483198424956552] 
\draw    (300,34) -- (300,268) ;
\draw [shift={(300,271)}, rotate = 270] [fill={rgb, 255:red, 0; green, 0; blue, 0 }  ][line width=0.08]  [draw opacity=0] (8.93,-4.29) -- (0,0) -- (8.93,4.29) -- cycle    ;
\draw [shift={(300,31)}, rotate = 90] [fill={rgb, 255:red, 0; green, 0; blue, 0 }  ][line width=0.08]  [draw opacity=0] (8.93,-4.29) -- (0,0) -- (8.93,4.29) -- cycle    ;
%Straight Lines [id:da5017839204505935] 
\draw    (397.43,209.46) -- (202.57,92.54) ;
\draw [shift={(200,91)}, rotate = 30.96] [fill={rgb, 255:red, 0; green, 0; blue, 0 }  ][line width=0.08]  [draw opacity=0] (8.93,-4.29) -- (0,0) -- (8.93,4.29) -- cycle    ;
\draw [shift={(400,211)}, rotate = 210.96] [fill={rgb, 255:red, 0; green, 0; blue, 0 }  ][line width=0.08]  [draw opacity=0] (8.93,-4.29) -- (0,0) -- (8.93,4.29) -- cycle    ;
%Straight Lines [id:da7732539643972135] 
\draw    (397.43,92.54) -- (202.57,209.46) ;
\draw [shift={(200,211)}, rotate = 329.04] [fill={rgb, 255:red, 0; green, 0; blue, 0 }  ][line width=0.08]  [draw opacity=0] (8.93,-4.29) -- (0,0) -- (8.93,4.29) -- cycle    ;
\draw [shift={(400,91)}, rotate = 149.04] [fill={rgb, 255:red, 0; green, 0; blue, 0 }  ][line width=0.08]  [draw opacity=0] (8.93,-4.29) -- (0,0) -- (8.93,4.29) -- cycle    ;
%Straight Lines [id:da5407139353956435] 
\draw    (328.46,103.57) -- (271.54,198.43) ;
\draw [shift={(270,201)}, rotate = 300.96] [fill={rgb, 255:red, 0; green, 0; blue, 0 }  ][line width=0.08]  [draw opacity=0] (8.93,-4.29) -- (0,0) -- (8.93,4.29) -- cycle    ;
\draw [shift={(330,101)}, rotate = 120.96] [fill={rgb, 255:red, 0; green, 0; blue, 0 }  ][line width=0.08]  [draw opacity=0] (8.93,-4.29) -- (0,0) -- (8.93,4.29) -- cycle    ;
%Straight Lines [id:da38320860998788053] 
\draw    (271.54,103.57) -- (328.46,198.43) ;
\draw [shift={(330,201)}, rotate = 239.04] [fill={rgb, 255:red, 0; green, 0; blue, 0 }  ][line width=0.08]  [draw opacity=0] (8.93,-4.29) -- (0,0) -- (8.93,4.29) -- cycle    ;
\draw [shift={(270,101)}, rotate = 59.04] [fill={rgb, 255:red, 0; green, 0; blue, 0 }  ][line width=0.08]  [draw opacity=0] (8.93,-4.29) -- (0,0) -- (8.93,4.29) -- cycle    ;
%Straight Lines [id:da49729778996070895] 
\draw    (243,151) -- (357,151) ;
\draw [shift={(360,151)}, rotate = 180] [fill={rgb, 255:red, 0; green, 0; blue, 0 }  ][line width=0.08]  [draw opacity=0] (8.93,-4.29) -- (0,0) -- (8.93,4.29) -- cycle    ;
\draw [shift={(240,151)}, rotate = 0] [fill={rgb, 255:red, 0; green, 0; blue, 0 }  ][line width=0.08]  [draw opacity=0] (8.93,-4.29) -- (0,0) -- (8.93,4.29) -- cycle    ;

% Text Node
\draw (180,73.4) node [anchor=north west][inner sep=0.75pt]    {$\vec{\alpha }_{1}$};
% Text Node
\draw (330,82.4) node [anchor=north west][inner sep=0.75pt]    {$\vec{\alpha }_{4}$};
% Text Node
\draw (260,76.4) node [anchor=north west][inner sep=0.75pt]    {$\vec{\alpha }_{3}$};
% Text Node
\draw (361,136.4) node [anchor=north west][inner sep=0.75pt]    {$\vec{\alpha }_{2}$};
% Text Node
\draw (400,72.4) node [anchor=north west][inner sep=0.75pt]    {$\vec{\alpha }_{5}$};
% Text Node
\draw (291,6.4) node [anchor=north west][inner sep=0.75pt]    {$\vec{\alpha }_{6}$};

\end{tikzpicture}
    \caption{Root system of $\mathfrak g_{2(2)}$.}
    \label{fig:roots}
\end{figure}

We begin by reviewing the mathematical structure of $G_{2(2)}$, following~\cite{Compere:2009zh}. Note that this is the split real form of $G_2$. The corresponding Lie algebra $\mathfrak g_{2(2)}$ is an exceptional Lie algebra of rank 2 and dimension 14, whose Dynkin diagram is given in Figure~\ref{fig:dynkin}. 
We denote the Cartan subalgebra by $\mathfrak h$ and the Cartan generators by $h_a$, $a=1,2$. There are then six positive generators $e_i$ and six corresponding negative generators $f_i$ normalized such that
\begin{equation}
    [\vec h,e_i]=\vec\alpha_i e_i,\qquad [\vec h,f_i]=-\vec\alpha_i f_i,\qquad [e_i,f_i]=\vec\alpha_i\cdot\vec h,\label{eq:commies}
\end{equation}
where $\vec h=(h_1,h_2)$ and where the positive roots are given by
\begin{align}
    \vec\alpha_1&=\qty(-\sqrt{3},1),&\vec\alpha_2&=\qty(\frac{2}{\sqrt{3}},0),\nonumber\\
    \vec\alpha_3&=\qty(-\frac{1}{\sqrt{3}},1)=\vec\alpha_1+\vec\alpha_2,&\vec\alpha_4&=\qty(\frac{1}{\sqrt{3}},1)=\vec\alpha_1+2\vec\alpha_2,\nonumber\\
    \vec\alpha_5&=\qty(\sqrt{3},1)=\vec\alpha_1+3\vec\alpha_2,&\vec\alpha_6&=\qty(0,2)=2\vec\alpha_1+3\vec\alpha_2.\label{eq:roots}
\end{align}
The root system is shown in Figure~\ref{fig:roots}. In this parametrization, $\vec\alpha_1$ and $\vec\alpha_2$ are the simple roots. Each node of the Dynkin diagram (Fig.~\ref{fig:dynkin}) corresponds to a triple of Chevalley generators $\{e_a,f_a,\vec\alpha_a\cdot\vec h\}$, $a=1,2$. Note that the $e_i$ have nontrivial commutation relations among themselves, which read
\begin{equation}
    [e_1,e_2]=e_3,\qquad [e_3,e_2]=e_4,\qquad [e_4,e_2]=e_5,\qquad [e_1,e_5]=e_6.
\end{equation}

The $e_i$ naturally span a nilpotent subalgebra $\mathfrak n_+$, while the $f_i$ span another nilpotent subalgebra $\mathfrak n_-$. However, we will define
\begin{align}
    k_1&=e_1+f_1,& k_2&=e_2+f_2,&k_3&=e_3-f_3,\nonumber\\
    k_4&=e_4+f_4,& k_5&=e_5-f_5,&k_6&=e_6+f_6.\label{eq:kdefs}
\end{align}
The $k_i$ then span a pseudocompact subalgebra $\mathfrak k$, which can be shown to be equivalent to $\mathfrak{sl}(2,\mathbb R)\oplus\mathfrak{sl}(2,\mathbb R)$. Note that~\eqref{eq:commies} and~\eqref{eq:kdefs} together imply that
\begin{equation}
    [e_i,k_i]=\vec\alpha_i\cdot \vec h.\label{eq:ekcomm}
\end{equation}
We may then decompose $\mathfrak g_{2(2)}$ as
\begin{equation}
    \mathfrak g_{2(2)}=\mathfrak h\oplus\mathfrak n_+\oplus\mathfrak k.
\end{equation}

\subsection{Two-derivative $G_{2(2)}$ symmetry}
We now proceed to review the $G_{2(2)}$ symmetry that arises upon reducing five-dimensional minimal supergravity to three dimensions. We start with minimal supergravity in five dimensions, which has a single symplectic Majorana supercharge. The field content is then just the $\mathcal N = 2$ gravity multiplet consisting of a metric $\hat g_{\hat\mu\hat\nu}$, a symplectic Majorana gravitino $\psi_{\hat\mu}$, and a graviphoton $\hat A_{\hat\mu}$. The two-derivative bosonic Lagrangian is given by
\begin{equation}
    \hat e^{-1}\mathcal L_5=\hat R-\frac{1}{4}\hat F_{\hat\mu\hat\nu}\hat F^{\hat\mu\hat\nu}+\frac{1}{12\sqrt{3}}\epsilon^{\hat\mu\hat\nu\hat\rho\hat\sigma\hat\lambda}\hat F_{\hat\mu\hat\nu}\hat F_{\hat\rho\hat\sigma}\hat A_{\hat\lambda},\label{eq:5daction}
\end{equation}
where $\hat R$ is the five-dimensional Ricci scalar and $\hat F_{(2)} = \dd\hat A_{(1)}$. Note that there is a trombone symmetry
\begin{equation}
    \hat g_{\hat\mu\hat\nu}\to\Lambda^2\hat g_{\hat\mu\hat\nu},\qquad\hat A_{(1)}\to\Lambda\hat A_{(1)},\qquad\Lambda\in\mathbb R^+,\label{eq:trombone}
\end{equation}
which rescales the action by an overall factor of $\Lambda^3$, while leaving the equations of motion invariant.

We reduce the five-dimensional fields \eqref{eq:5daction} using the ansatz
\begin{align}
    \dd s_5^2&=\tau^{-1}g_{\mu\nu}\dd x^\mu\dd x^\nu+g_{ij}\qty(\dd y^i+\omega^i_\mu\dd x^\mu)\qty(\dd y^j+\omega^j_\nu\dd x^\nu),\nonumber\\
    \hat A_{(1)}&=\bar A_\mu\dd x^\mu+a_i\qty(\dd y^i+\omega^i_\mu\dd x^\mu),\label{eq:redAnsatz}
\end{align}
where $y^i=\{z,t\}$ are coordinates on the internal space,\footnote{We are focusing on the case of reducing along one timelike and one spacelike direction, which is not technically a torus. However, by abuse of language, we will refer to this as $T^2$.} $x^\mu$ are coordinates on the three-dimensional base space, and $\tau=|\det g_{ij}|$. Here, $g_{\mu\nu}$ functions as a metric on the base space, $\omega^i_\mu$ as a $U(1)^2$ gauge field, and $g_{ij}$ as a symmetric matrix of scalars. Plugging our ansatz~\eqref{eq:redAnsatz} into our action~\eqref{eq:5daction} yields the three-dimensional Lagrangian
\begin{align}
    e^{-1}\mathcal L_3=R-\frac{1}{2}\frac{(\partial\tau)^2}{\tau^2}-\frac{1}{4}\tau \bar F^2-\frac{1}{4}\tau\, g_{ij}\bar W^i_{\mu\nu}\bar W^{j\mu\nu}-\frac{1}{4}g^{ij}g^{kl}\partial_\mu g_{ik}\partial^\mu g_{jl}-\frac{1}{2}g^{ij}\partial_\mu a_i\partial^\mu a_j+\mathrm{CS},%+\frac{2}{\sqrt{3}}\epsilon^{\mu\nu\rho}\partial_\mu a_1\partial_\nu a_2\bar A_\rho.
    \label{eq:3dActFirstForm}
\end{align}
where $\bar F_{(2)}=\dd\bar A_{(1)}$ and $W^i_{(2)}=\dd\omega^i_{(1)}$, and $\mathrm{CS}$ schematically denotes the CP-odd terms that we will be more precise about later.

Note that the five-dimensional local $GL(5,\mathbb R)$ diffeomorphism symmetry has split into a three-dimensional local $GL(3,\mathbb R)$ diffeomorphism symmetry, a local $U(1)^2$ gauge symmetry, and a global $GL(2,\mathbb R)$ of large diffeomorphisms on the torus.\footnote{Strictly speaking, large diffeomorphisms on $T^2$ constitute $SL(2,\mathbb Z)$. However, when we truncate to the zero-mode sector in the classical Kaluza-Klein reduction, this is enhanced to a $GL(2,\mathbb R)$.} This consists of an $SL(2,\mathbb R)$ subgroup that corresponds to volume-preserving reparametrizations of the torus,
\begin{equation}
    y^i\to\lambda^i{}_j y^j,\qquad g_{ij}\to\lambda_i{}^k\lambda_j{}^lg_{kl},\qquad \omega_{(1)}^i\to\lambda^i{}_j\omega^j_{(1)},\qquad a_i\to \lambda_i{}^ja_j,\qquad \lambda\in SL(2,\mathbb R),\label{eq:sl2trans}
\end{equation}
which leave the action~\eqref{eq:3dActFirstForm} invariant, along with an $\mathbb R^+$ subgroup that rescales the volume of the torus as
\begin{equation}
    y^i\to\Lambda y^i,\qquad g_{ij}\to\Lambda^2g_{ij},\qquad\omega_{(1)}^i\to\Lambda^{-1}\omega_{(1)}^i,\qquad a_i\to\Lambda a_i,\qquad g_{\mu\nu}\to\Lambda^{4}g_{\mu\nu},\qquad\Lambda\in\mathbb R^+,\label{eq:R+scale}
\end{equation}
which rescales the action~\eqref{eq:3dActFirstForm} by an overall factor $\mathcal L_3\to\Lambda^{2}\mathcal L_3$, but leaves the equations of motion invariant. 

Note that there are also $SO(2)$ local Lorentz transformations of the torus frame bundle, which means that the resulting Kaluza-Klein theory may be thought of as a $GL(2,\mathbb R)/SO(2)$ coset model. If we write $g_{ij}=e_i^a e_j^b\eta_{ab}$ in terms of a torus vielbein $e_i^a$, then $e_i^a$ forms a coset representative of $GL(2,\mathbb R)/SO(2)$, transforming as
\begin{equation}
    e_i^a\to \sigma^a{}_b\, e_j^b\, \lambda^j{}_i,\qquad \lambda\in GL(2,\mathbb R),\quad \sigma\in SO(2).
\end{equation}
Note that $GL(2,\mathbb R)$ acts globally on $e_i^a$, whereas $SO(2)$ acts locally.

We note that there is also the five-dimensional trombone symmetry~\eqref{eq:trombone}, which may be expressed in terms of three-dimensional fields as
\begin{equation}
    g_{\mu\nu}\to\Lambda^6 g_{\mu\nu},\qquad \bar A_{(1)}\to\Lambda\bar A_{(1)},\qquad g_{ij}\to\Lambda^2 g_{ij},\qquad a_i\to\Lambda a_i,\qquad\Lambda\in\mathbb R^+,\label{eq:trombone3D}
\end{equation}
which rescales the two-derivative action by a homogeneous factor $\mathcal L_3\to\Lambda^3\mathcal L_3$. Combining this with the scaling~\eqref{eq:R+scale}, one finds an $\mathbb R^+$ scaling transformation that leaves the metric invariant,
\begin{equation}
    \bar A_{(1)}\to\Lambda^{-2}\bar A_{(1)},\qquad\omega^i_{(1)}\to\Lambda^{-3}\omega^i_{(1)},\qquad g_{ij}\to\Lambda^2 g_{ij},\qquad a_i\to\Lambda a_i,\qquad\Lambda\in\mathbb R^+.\label{eq:R+combo}
\end{equation}
We remark that the $GL(2,\mathbb R)$ symmetry came purely from geometry, whereas the scaling symmetry~\eqref{eq:R+combo} is deeply related to the presence of the five-dimensional trombone symmetry. This will be important for higher-derivative corrections.

In order to make the $G_{2(2)}$ symmetry manifest, we will now switch to the parametrization~\cite{Compere:2009zh}
\begin{align}
    g_{ij}&=\begin{pmatrix}e^{-\tfrac{2}{\sqrt{3}}\phi_1}&e^{-\tfrac{2}{\sqrt{3}}\phi_1}\chi_1\\
    e^{-\tfrac{2}{\sqrt{3}}\phi_1}\chi_1&\quad-e^{\tfrac{1}{\sqrt{3}}\phi_1-\phi_2}+e^{-\tfrac{2}{\sqrt{3}}\phi_1}\chi_1^2\end{pmatrix},\nonumber\\
    \tau&=e^{-\tfrac{1}{\sqrt{3}}\phi_1-\phi_2},\nn\\
    \omega_{(1)}^1&=\mathcal A^1_{(1)}-\chi_1\mathcal A^2_{(1)},\qquad\omega_{(1)}^2=\mathcal A^2_{(1)},\nonumber\\
    a_1&=\chi_2,\qquad a_2=\chi_3,\nonumber\\
    A_{(1)}&=\bar A_{(1)}+a_i\omega^i_{(1)}.
\end{align}
Note that this is simply a redefinition of fields in our reduction ansatz. We then define invariant field strengths for the gauge fields
\begin{align}
    \mathcal F^1_{(2)}&=\dd\mathcal A^1_{(1)}+\mathcal A^2_{(1)}\land\dd\chi_1,\nonumber\\
    \mathcal F^2_{(2)}&=\dd\mathcal A^2_{(1)},\nonumber\\
    F_{(2)}&=\dd A_{(1)}-\dd\chi_2\land\qty(\mathcal A^1_{(1)}-\chi_1\mathcal A^2_{(1)})-\dd\chi_3\land\mathcal A^2_{(1)},
\end{align}
and for the scalars
\begin{align}
    F_{(1)}^1&=\dd\chi_1,\nonumber\\
    F_{(1)}^2&=\dd\chi_2,\nonumber\\
    F_{(1)}^3&=\dd\chi_3-\chi_1\,\dd\chi_2.
\end{align}
The reduced three-dimensional Lagrangian is then given by~\cite{Cremmer:1998em,Cremmer:1999du}
\begin{align}
    e^{-1}\mathcal L_3&=R-\frac{1}{2}\partial\vec\phi\cdot\partial\vec\phi+\frac{1}{2}e^{\vec\alpha_1\cdot\vec\phi}\qty(F^1)^2+\frac{1}{2}e^{\vec\alpha_2\cdot\vec\phi}\qty(F^2)^2-\frac{1}{2}e^{\vec\alpha_3\cdot\vec\phi}\qty(F^3)^2\nonumber\\
    &\qquad-\frac{1}{4}e^{-\vec\alpha_4\cdot\vec\phi}F^2+\frac{1}{4}e^{-\vec\alpha_5\cdot\vec\phi}\qty(\mathcal F^1)^2-\frac{1}{4}e^{-\vec\alpha_6\cdot\vec\phi}\qty(\mathcal F^2)^2+\frac{2}{\sqrt{3}}\epsilon^{\mu\nu\rho}F^2_\mu F^3_\nu A_\rho,\label{eq:3dAct}
\end{align}
where $\vec\phi=(\phi_1,\phi_2)$. Note that the $\vec\alpha_i$ in~\eqref{eq:3dAct} correspond precisely to the positive roots~\eqref{eq:roots} of $\mathfrak g_{2(2)}$, which is the first sign of $G_{2(2)}$ structure.

Since we are working in three dimensions, vectors are Hodge dual to scalars, which we exploit by dualizing~\cite{Compere:2009zh}
\begin{align}
    e^{-\vec\alpha_4\cdot\vec\phi}F_{(2)}&\equiv G_{(1)\,4}=\dd\chi_4+\frac{1}{\sqrt{3}}\qty(\chi_2\dd\chi_3-\chi_3\dd\chi_2),\nonumber\\
    e^{-\vec\alpha_5\cdot\vec\phi}\mathcal F^1_{(2)}&\equiv G_{(1)\,5}=\dd\chi_5-\chi_2\dd\chi_4+\frac{\chi_2}{3\sqrt{3}}\qty(\chi_3\dd\chi_2-\chi_2\dd\chi_3),\nonumber\\
    e^{-\vec\alpha_6\cdot\vec\phi}\mathcal F^2_{(2)}&\equiv G_{(1)\,6}=\dd\chi_6-\chi_1\dd\chi_5+\qty(\chi_1\chi_2-\chi_3)\dd\chi_4+\frac{1}{3\sqrt{3}}\qty(\chi_1\chi_2-\chi_3)\qty(\chi_2\dd\chi_3-\chi_3\dd\chi_2).\label{eq:hodge}
\end{align}
Note that dualization exchanges the equations of motion and Bianchi identities for the fields and their duals, and so the Chern-Simons term in the action~\eqref{eq:3dAct} leads to the above parametrization of the $G_{(1)\,i}$ in terms of $\chi_i$. With this, the three-dimensional action then takes the form~\cite{Compere:2009zh}
\begin{align}
    e^{-1}\mathcal L_3&=R-\frac{1}{2}\partial\vec\phi\cdot\partial\vec\phi+\frac{1}{2}e^{\vec\alpha_1\cdot\vec\phi}\qty(F^1)^2+\frac{1}{2}e^{\vec\alpha_2\cdot\vec\phi}\qty(F^2)^2-\frac{1}{2}e^{\vec\alpha_3\cdot\vec\phi}\qty(F^3)^2\nonumber\\
    &\kern7em\ +\frac{1}{2}e^{\vec\alpha_4\cdot\vec\phi}\qty(G_4)^2-\frac{1}{2}e^{\vec\alpha_5\cdot\vec\phi}\qty(G_5)^2+\frac{1}{2}e^{\vec\alpha_6\cdot\vec\phi}\qty(G_6)^2.
\end{align}
In particular, the action now takes the form of gravity coupled to eight scalar fields, namely $\vec\phi$ and $\chi_i$. Note that the unusual signs of the scalar kinetic terms are due to reducing over a timelike direction.

To show the manifest invariance of the action under $G_{2(2)}$, we should write it in terms of covariant objects. Exponentiating the generators of $\mathfrak g_{2(2)}$ with scalars as coefficients allows one to construct
\begin{equation}
    \mathcal V=e^{\tfrac{1}{2}\phi_1h_1+\tfrac{1}{2}\phi_2h_2}e^{\chi_1e_1}e^{-\chi_2e_2+\chi_3e_3}e^{\chi_6e_6}e^{\chi_4e_4-\chi_5e_5},
\end{equation}
as a coset representative of $G_{2(2)}/K$, where $K=SL(2,\mathbb R)\times SL(2,\mathbb R)$ is the maximal pseudocompact subgroup obtained by exponentiating $\mathfrak k$. Note that this corresponds to a particular choice of gauge~\cite{Keurentjes:2005jw,Bossard:2009at}. This representative transforms as
\begin{equation}
    \mathcal V\to k\mathcal Vg,\qquad k\in K,\quad g\in G_{2(2)}.
\end{equation}
Note that here $G_{2(2)}$ acts globally on $\mathcal V$, whereas $K$ acts locally. One then defines a scalar matrix
\begin{equation}
    \mathcal M=\mathcal V^\#\mathcal V,
\end{equation}
where $\#$ denotes generalized transposition, which is an involution defined on the generators of $\mathfrak g_{2(2)}$ by
\begin{align}
    \#(h_a)&=h_a,\nonumber\\
    \#(e_1)&=-f_1,\qquad\#(e_2)=-f_2,\qquad\#(e_3)=f_3,\nonumber\\
    \#(e_4)&=-f_4,\qquad\#(e_5)=f_5,\qquad\ \ \#(e_6)=-f_6.
\end{align}
The scalar matrix transforms covariantly under global $G_{2(2)}$ transformations
\begin{equation}
    \mathcal M\to g^\#\mathcal M g,\qquad g\in G_{2(2)}.
\end{equation}
As such, the action may be rewritten as a non-linear sigma model~\cite{Compere:2009zh}
\begin{align}
    e^{-1}\mathcal L_3=R-\frac{1}{8}\Tr\qty(\mathcal M^{-1}\partial_\mu\mathcal M\mathcal M^{-1}\partial^\mu\mathcal M),
\end{align}
where the trace is over $G_{2(2)}$ indices. Thus, we see a symmetry enhancement from the geometric $GL(2,\mathbb R)$ to $G_{2(2)}$.

Now, let us make several comments regarding the action of $G_{2(2)}$. First, note that $\beta k_1$ exponentiates to a boost along the $t$-$z$ plane of the torus~\cite{Compere:2009zh}
\begin{equation}
    \begin{pmatrix}
        t\\z
    \end{pmatrix}\to\begin{pmatrix}
        \cosh\beta&\sinh\beta\\\sinh\beta&\cosh\beta
    \end{pmatrix}\begin{pmatrix}
        t\\z
    \end{pmatrix},
\end{equation}
while $\beta e_1$ exponentiates to a volume-preserving rescaling of the torus
\begin{equation}
    t\to\beta t,\qquad z\to\beta^{-1}z.
\end{equation}
Together with $\vec\alpha_1\cdot\vec h$, these form the volume-preserving $SL(2,\mathbb R)$ that we saw earlier in~\eqref{eq:sl2trans}. Thus, the generators of $\mathfrak{sl}(2,\mathbb R)=\langle e_1,k_1,\vec\alpha_1\cdot\vec h\rangle$ precisely correspond to the triple of Chevalley generators associated to the first node of the Dynkin diagram of $\mathfrak{g}_{2(2)}$ (Fig.~\ref{fig:dynkin}). Note that neither the $\mathbb R^+\subset GL(2,\mathbb R)$ scaling~\eqref{eq:R+scale} nor the trombone symmetry~\eqref{eq:trombone3D} directly corresponds to a $G_{2(2)}$ transformation, as they cannot be written solely in terms of an action on the moduli. However, the other linearly independent Cartan generator $\vec\alpha_4\cdot\vec h$ corresponds to the $\mathbb R^+$ scaling transformation~\eqref{eq:R+combo}, which arises from the combination of the $GL(2,\mathbb R)$ scaling and the trombone symmetry.

The $e_i$ correspond to (large) gauge transformations. In particular, $b_1e_1+\cdots+b_6e_6$ act on the first three $\chi_i$ as electric gauge transformations~\cite{Compere:2009zh}
\begin{align}
    \chi_1&\to\chi_1+b_1,\nonumber\\
    \chi_2&\to\chi_2-b_2,\nonumber\\
    \chi_3&\to\chi_3+b_1\chi_2-\frac{1}{2}b_1b_2+b_3.
\end{align}
It is straightforward to see that these leave $\mathcal F_{(1)}$, $F^1_{(1)}$, and $F^2_{(1)}$ invariant. The transformations of $\chi_4$, $\chi_5$, and $\chi_6$ are somewhat more complicated (and nonlinear) but correspond to magnetic gauge transformations that leave the $G_{(1)\, i}$ invariant. We remark that the $e_1$ symmetry arises from rigid diffeomorphisms of the torus, the $\langle e_2, e_3\rangle$ symmetry comes from the residual (large) gauge invariance of the five-dimensional gauge field along the torus directions
\begin{equation}
    \hat A\to\hat A-b_2\dd z+b_3\dd t,
\end{equation}
and the $\langle e_4, e_5,e_6\rangle=\mathbb R^3$ symmetry algebra comes from the magnetic (large) gauge transformation duals of the original $U(1)^3$ gauge symmetry.

The $k_i$ span the algebra $\mathfrak k=\mathfrak{sl}(2,\mathbb R)\oplus\mathfrak{sl}(2,\mathbb R)$, which can be used to generate non-trivial solutions~\cite{Bouchareb:2007ax,Tomizawa:2008qr,Compere:2009zh,Suzuki:2024coe,Suzuki:2024phv,Suzuki:2024vzq}. Aside from $k_1$, these are non-geometric in origin and correspond to Harrison-Ehlers and $S$-duality transformations.

\subsection{Four-derivative $G_{2(2)}$ symmetry}
There is a unique supersymmetric four-derivative extension of minimal five-dimensional supergravity, given by~\cite{Liu:2022sew}
\begin{equation}
    \hat e^{-1}\mathcal L_{4\partial}=\hat{\mathcal{X}}_4-\frac{1}{2}\hat C_{\hat\mu\hat\nu\hat\rho\hat\sigma}\hat F^{\hat\mu\hat\nu}\hat F^{\hat\rho\hat\sigma}+\frac{1}{8}\hat F^4+\frac{1}{2\sqrt{3}}\epsilon^{\hat\mu\hat\nu\hat\rho\hat\sigma\hat\lambda}\hat R_{\hat\mu\hat\nu\hat\alpha\hat\beta}\hat R_{\hat\rho\hat\sigma}{}^{\hat\alpha\hat\beta}\hat A_{\hat\lambda},\label{eq:4derAct}
\end{equation}
where $\hat C_{\hat\mu\hat\nu\hat\rho\hat\sigma}$ is the five-dimensional Weyl tensor and $\hat{\mathcal{X}}_4=\hat R_{\hat\mu\hat\nu\hat\rho\hat\sigma}\hat R^{\hat\mu\hat\nu\hat\rho\hat\sigma}-4\hat R_{\hat\mu\hat\nu}\hat R^{\hat\mu\hat\nu}+\hat R^2$ is the five-dimensional Gauss-Bonnet combination. In principle, the brute-force computation would be to reduce this action, field redefine as appropriate, and then try to rearrange the terms into $G_{2(2)}$ invariant combinations. However, there is a simpler argument from group theory.

First, note that the $\mathbb R^+$ symmetry is generically broken by higher-derivative corrections. For example, $\sqrt{-\hat g}\hat R_{\hat\mu\hat\nu\hat\rho\hat\sigma}^2$ will reduce to $\sqrt{g}\tau R_{\mu\nu\rho\sigma}^2+\cdots$,
$\sqrt{-\hat g}\hat F^4$ will reduce to $\sqrt{g}\tau^3F^4+\cdots$, \emph{etc}. Such terms then break the $\mathbb R^+$ scaling symmetry~\eqref{eq:R+combo}. This is a generic feature of higher-derivative corrections: Even taking the most general possible four-derivative corrections up to field redefinitions, there is no choice of coefficients that would allow one to preserve the $\mathbb R^+$ scaling symmetry. This is a consequence of dimensional analysis as the two-derivative Lagrangian $\mathcal L_5$ is a dimension-five operator, while the four-derivative Lagrangian $\mathcal L_{4\partial}$ is a dimension-seven operator. This is directly due to the fact that~\eqref{eq:4derAct} breaks the trombone symmetry~\eqref{eq:trombone}, since it scales homogeneously as $\mathcal L_{4\partial}\to\Lambda\mathcal L_{4\partial}$. Nevertheless, the volume-preserving large diffeomorphisms of the torus corresponding to the global $\mathfrak{sl}(2,\mathbb R)=\langle e_1,k_1,\vec\alpha_1\cdot\vec h\rangle$ will be preserved, as these originate from five-dimensional diffeomorphism invariance. Similarly, we expect the large gauge symmetries $e_i$ to be preserved, as they arise from the gauge and diffeomorphism invariance of the five-dimensional parent theory.

As such, the four-derivative corrections~\eqref{eq:4derAct} will explicitly break the symmetries from $\mathfrak g_{2(2)}$ to some proper subalgebra $\mathfrak l$ that contains $\vec\alpha_1\cdot\vec h$, $k_1$, and all the $e_i$, but does not contain $\vec\alpha_4\cdot\vec h$. Demanding closure of the algebra under the Lie bracket, we immediately see from the commutation relation~\eqref{eq:ekcomm} and the expressions for the roots~\eqref{eq:roots} that all $k_i$ (except for $k_1$) must be broken. That is,
\begin{equation}
    \mathfrak l=\{\vec\alpha_1\cdot\vec h\}\oplus\mathfrak n_+\oplus \{k_1\}.\label{eq:residAlg}
\end{equation}
This corresponds to preserving only the electric/magnetic gauge transformations and the volume-preserving large diffeomorphisms of the torus, which are the geometric symmetries for any Kaluza-Klein theory on a torus. Said more bluntly, the four-derivative corrections ruin all symmetry enhancement that was present at the two-derivative level.

\let\oldaddcontentsline\addcontentsline% Store \addcontentsline
\renewcommand{\addcontentsline}[3]{}% Make \addcontentsline a no-op
\subsection*{Geometric subalgebra}
As an aside, one can ask to which Lie algebra~\eqref{eq:residAlg} corresponds. One finds that ${\langle e_2,\cdots,e_6\rangle}$ constitutes the radical. Fortunately, five-dimensional solvable Lie algebras have been classified, and we see that this corresponds to $\mathfrak g_{5,4}$ in the classification of~\cite{Dixmier:1958}. We then get a short exact sequence
\begin{equation}
    0\to \mathfrak g_{5,4}\to\mathfrak l\to\mathfrak l/\mathfrak g_{5,4}\to 0.
\end{equation}
The semisimple algebra $\mathfrak l/\mathfrak g_{5,4}$ is spanned by $\langle\vec\alpha_1\cdot\vec h,e_1,k_1\rangle$ and precisely corresponds to the $\mathfrak{sl}(2,\mathbb R)$ volume-preserving diffeomorphisms. Thus, upon exponentiation, we find that the symmetries of the four-derivative theory are $SL(2,\mathbb R)\ltimes L_{5,4}$, where $\mathrm{Lie}\qty(L_{5,4})=\mathfrak g_{5,4}$. This is not so surprising as the breaking of $\vec\alpha_4\cdot\vec h$ means that $\vec\alpha_2\cdot\vec h$ will be broken, and, as such, corresponds to deletion of the second node of the Dynkin diagram of $\mathfrak{g}_{2(2)}$, leaving a single Chevalley triple corresponding to $\mathfrak{sl}(2,\mathbb R)$.

\let\addcontentsline\oldaddcontentsline% Restore \addcontentsline

\subsection{$SL(3,\mathbb R)$ symmetry in pure gravity}\label{sec:SL3}
There is a special case of the above analysis that applies to pure gravity. That is, the Einstein-Hilbert action,
\begin{equation}
    \hat e^{-1}\mathcal L_5=\hat R,
\end{equation}
reduced on $T^2$ leads to an $SL(3,\mathbb R)$ symmetry~\cite{Maison:1979kx,Maison:2000fj}. This can be seen as a consistent truncation of the minimal supergravity case where we set $\hat A=0$, which corresponds in the three-dimensional description to truncating
\begin{equation}
    A_\mu=0,\qquad \chi_2=0,\qquad\chi_3=0.
\end{equation}
This truncates the field strengths
\begin{equation}
    F_{(2)}=0,\qquad F^2_{(1)}=0,\qquad F^3_{(1)}=0.\label{eq:FStrunc}
\end{equation}
After Hodge dualization~\eqref{eq:hodge}, the first condition of~\eqref{eq:FStrunc} is equivalent to
\begin{equation}
    G_{(1)\,4}=0.
\end{equation}
As such, the action of $e_2$, $e_3$, and $e_4$ on the moduli becomes trivial, as do, correspondingly, $f_2$, $f_3$, and $f_4$. Consequently, the remaining generators of the algebra are $h_1$, $h_2$, $e_1$, $e_5$, $e_6$, $f_1$, $f_5$, $f_6$, which can be seen to span $\mathfrak{sl}(3,\mathbb R)$. In particular, the geometric $\mathfrak{sl}(2,\mathbb R)=\langle \vec\alpha_1\cdot\vec h,e_1,k_1\rangle$ is still present, as we would expect. Incidentally, the radical of $\langle\vec\alpha_1\cdot\vec h,e_1,e_5,e_6, k_1\rangle$ is given by $\langle e_5,e_6\rangle=\mathbb R^2$, and so the geometric global symmetries are given by $SL(2,\mathbb R)\ltimes U(1)^2$, which is enhanced by the hidden symmetry to the full $SL(3,\mathbb R)$.

In principle, the higher-derivative failure of $SL(3,\mathbb R)$ symmetry enhancement is automatically implied by the failure of $G_{2(2)}$ symmetry enhancement. However, due to the $\Tr\qty(\hat R\land\hat R)\land \hat A$ term in the four-derivative action, this will not be a consistent truncation. Nevertheless, the only four-derivative term we may write down (up to field redefinitions) is
\begin{equation}
    \hat e^{-1}\mathcal L_{4\partial}=\hat R_{\hat\mu\hat\nu\hat\rho\hat\sigma}\hat R^{\hat\mu\hat\nu\hat\rho\hat\sigma},
\end{equation}
which, as before, will not preserve the $\mathbb R^+$ scaling symmetry~\eqref{eq:R+combo} upon dimensional reduction. As such, an identical group theory argument applies as for the $G_{2(2)}$ case. In particular, we expect that $\langle\vec\alpha_1\cdot\vec h,e_1,e_5,e_6, k_1\rangle$ is preserved and $\vec\alpha_4\cdot\vec h$ is broken. Closure of the commutation relations then implies that $k_5$ and $k_6$ must be broken. The key point is that our argument does not rely on the specific form of the higher-derivative action, only that it generically breaks the scaling symmetry.

\section{$O(d+p+1,d+1)$ U-duality}\label{sec:Odpd}
We now consider the U-duality of heterotic/bosonic supergravity reduced to three dimensions. We start with the two-derivative supergravity action in $d+3$ dimensions
\begin{equation}
    \hat e^{-1}\mathcal L_{d+3}=e^{-2\phi}\qty(\hat R+4(\partial\phi)^2-\frac{1}{12}\hat H_{(3)}^2-\frac{1}{4}\Tr\hat{\mathcal F}^2_{(2)}),
    \label{eq:het0}
\end{equation}
where
\begin{equation}
    \hat H_{(3)}=\dd\hat B_{(2)}+\Tr\qty(\hat{\mathcal A}_{(1)}\land\hat{\mathcal F}_{(2)}),\qquad \hat{\mathcal F}_{(2)}=\dd\hat{\mathcal A}_{(1)}+\hat{\mathcal A}_{(1)}\land\hat{\mathcal A}_{(1)},
\end{equation}
where the trace is over the gauge group. As we will be interested in the class of backgrounds for which only the gauge fields corresponding to the Cartan generators of the gauge group take nontrivial values, we will henceforth assume that $\hat{\mathcal A}_{(1)}$ has gauge group $U(1)^p$ for simplicity. We will treat $d$ and $p$ as free parameters. There are four cases of interest to us, shown in Table~\ref{tab:cases}.
\begin{table}[h!]
    \centering
    \begin{tabular}{ || c | c | l || }
\hline
 ~$d$~ & ~$p$~ & ~Theory~ \\ 
 \hline
 ~23~ & ~0~ & ~Bosonic string~ \\  
 7 & ~0~ & ~Type II string$^*$~\\
 7&~$\le16$~~&~Heterotic string~~\\
 3&0&~STU model\\
 \hline
\end{tabular}
    \caption{Various physically relevant choices of $d$ and $p$.}
    \label{tab:cases}
\end{table}

\noindent Note that for the Type II strings, this corresponds to truncating the RR fields, which will give a proper subgroup of the full $E_{8(8)}$ U-duality group. In particular, the lack of an RR zero-form $C_{(0)}$ prevents the appearance of a modular function as in the IIB case~\eqref{eq:SL2invAction}, which makes this somewhat pathological. We reduce on $T^d$ using the standard ansatz
\begin{align}
    \dd s_{d+3}^2&=e^{4\varphi}g_{\mu\nu}\dd x^\mu\dd x^\nu+g_{ij}\eta^i\eta^j,\qquad\eta^i_{(1)}=\dd y^i+A^i_\mu\dd x^\mu,\nonumber\\
    \hat B_{(2)}&=\frac{1}{2}b_{\mu\nu}\dd x^\mu\land\dd x^\nu+B_{\mu i}\,\dd x^\mu\land\eta^i_{(1)}+\frac{1}{2}b_{ij}\,\eta^i_{(1)}\land\eta^j_{(1)},\nonumber\\
    \mathcal A^{\mathfrak a}_{(1)}&=\mathcal A^{\mathfrak a}_{(1)}\dd x^\mu+\mathcal A_i^{\mathfrak a}\eta^i_{(1)},\nonumber\\
    \phi&=\varphi+\frac{1}{4}\log|\det g_{ij}|.
\end{align}
As before, $y^i$ are the coordinates on the internal $T^d$ and $x^\mu$ are coordinates on the three-dimensional base space. $g_{\mu\nu}$ is interpreted as a three-dimensional metric; $A^i_{(1)}$, $B_{(1)\,i}$, and $\mathcal A^{\mathfrak a}_{(1)}$ as gauge fields; and $g_{ij}$, $b_{ij}$, and $\mathcal A^{\mathfrak a}_i$ as matrices of scalars. We will be agnostic about whether the internal space has a timelike direction. The gauge-invariant field strengths are given by
\begin{align}
    \hat H_{(3)}&=\frac{1}{3!}h_{\mu\nu\rho}\dd x^\mu\land\dd x^\nu\land\dd x^\rho+\frac{1}{2}G_{\mu\nu i}\dd x^\mu\land\dd x^\nu\land\eta^i_{(1)}+\frac{1}{2}F_{\mu ij}\dd x^\mu\land\eta^i_{(1)}\land\eta^j_{(1)},\nn\\
    \hat{\mathcal F}_{(2)}^{\mathfrak a}&=\mathcal F^{\mathfrak a}_{(2)}+\dd\mathcal A_i^{\mathfrak a}\land\eta^i_{(1)},
\end{align}
where
\begin{align}
    h_{(3)}&=\dd b_{(2)}+\frac{1}{2}\mathcal A_{(1)}^{\mathfrak a}\land\mathcal F^{\mathfrak a}_{(2)}-\frac{1}{2}B_{(1)\,i}\land F^i_{(2)},\nonumber\\
    G_{(2)\,i}&=\dd B_{(1)\,i}-b_{ij}F^{j}_{(2)}+\frac{1}{2}\mathcal A^{\mathfrak a}_{(1)}\land\dd\mathcal A_i^{\mathfrak a}+\frac{1}{2}\mathcal A_i^{\mathfrak a}\mathcal F_{(2)}^{\mathfrak a},\nonumber\\
    F_{(1)\,ij}&=\dd b_{ij}-\mathcal A_{[i}^{\mathfrak a}\dd\mathcal A_{j]}^{\mathfrak a},\nonumber\\
    \mathcal F_{(2)}^{\mathfrak a}&=\dd\mathcal A^{\mathfrak a}_{(1)}+\mathcal A_i^{\mathfrak a}F^i_{(2)},\nonumber\\
     F_{(2)}^i&=\dd A_{(1)}^i.
\end{align}
The resulting three-dimensional Lagrangian is then given by
\begin{align}
    e^{-1}\mathcal L_3&=R-4(\partial\varphi)^2-\frac{1}{12}e^{-8\varphi}h_{(3)}^2-\frac{1}{4}e^{-4\varphi}g_{ij}F^i_{\mu\nu}F^{j\,\mu\nu}-\frac{1}{4}e^{-4\varphi}g^{ij}G_{i\,\mu\nu}G_{j}^{\mu\nu}-\frac{1}{4}e^{-4\varphi}\mathcal F_{\mu\nu}^{\mathfrak a}\mathcal F^{\mathfrak a\mu\nu}\nonumber\\
    &\qquad\ \!-\frac{1}{4}g^{ij}g^{kl}\qty(\partial_\mu g_{ik}\partial^\mu g_{jl}+F_{\mu ik}F^\mu_{jl})-\frac{1}{2}g^{ij}\partial_\mu\mathcal A^{\mathfrak a}_i\partial^\mu\mathcal A^{\mathfrak a}_j.
\end{align}
As in the $G_{2(2)}$ case discussed earlier, there is a geometric $GL(d,\mathbb R)$ that arises from large diffeomorphisms of the torus, which acts on the three-dimensional fields as
\begin{align}
    g_{ij}&\to\lambda_i{}^k\lambda_j{}^lg_{kl},\quad A^i_{(1)}\to\lambda^i{}_jA^j_{(1)},\quad B_{(1)\,i}\to\lambda_i{}^jB_{(1)\,j},\quad b_{ij}\to\lambda_i{}^k\lambda_j{}^lb_{kl},\quad\mathcal A_i^{\mathfrak a}\to\lambda_i{}^j\mathcal A_j^{\mathfrak a},\nn\\
    y^i&\to\lambda^i{}_jy^j,\qquad \lambda\in GL(d,\mathbb R).
\end{align}
However, notice that the factor of $\tau=|\det g_{ij}|$ in the metric has been absorbed into the three-dimensional dilaton $\varphi$. In contrast to the $G_{2(2)}$ case, the nontrivial scaling of $\tau$ under $\mathbb R^+\subset GL(d,\mathbb R)$ is absorbed by a scaling of $\phi$. As such, the full $GL(d,\mathbb R)$ will have an action on the moduli, rather than just an $SL(d,\mathbb R)$.

If one defines
\begin{align}
    \mathcal H&=\begin{pmatrix}
        \quad g_{ij}+c_{ki}g^{kl}c_{lj}+\mathcal A_i^{\mathfrak a}\mathcal A_j^{\mathfrak a}&\quad g^{jk}c_{ki}&\quad - c_{ki}g^{kl}\mathcal A_l^{\mathfrak b}-\mathcal A_i^{\mathfrak b}\\
        g^{ik}c_{kj}&g^{ij}&\quad -g^{ik}\mathcal A_k^{\mathfrak b}\\
        - c_{kj}g^{kl}\mathcal A_l^{\mathfrak a}-\mathcal A_j^{\mathfrak a}&\quad -g^{jk}\mathcal A_k^{\mathfrak a}&\quad \delta_{\mathfrak a\mathfrak b}+\mathcal A_k^{\mathfrak a}g^{kl}\mathcal A^{\mathfrak a}_l
    \end{pmatrix},\nn\\
    \mathbb A^M_\mu&=\begin{pmatrix}
        A_\mu^i\\B_{\mu i}-\frac{1}{2}\mathcal A_\mu^{\mathfrak a}\mathcal A_i^{\mathfrak a}\\
        \mathcal A_\mu^{\mathfrak a}
    \end{pmatrix},\qquad
    \eta_{MN}=\begin{pmatrix}
        0&\quad\delta_i{}^{j}&\quad0\\
        \delta^i{}_{j}&\quad0&\quad0\\
        0&\quad0&\quad\delta_{\mathfrak a\mathfrak b}
    \end{pmatrix}.\label{eq:OdpdCoset}
\end{align}
where
\begin{equation}
    c_{ij}=b_{ij}-\frac{1}{2}\mathcal A_i^{\mathfrak a}\mathcal A_j^{\mathfrak a},
\end{equation}
then the three-dimensional action may be rewritten as
\begin{align}
    e^{-1}\mathcal L_3=R-4(\partial\varphi)^2+\frac{1}{8}\Tr\qty(\partial_\mu\mathcal H\partial^\mu\mathcal H^{-1})-\frac{1}{12}e^{-8\varphi}h_{\mu\nu\rho}h^{\mu\nu\rho}-\frac{1}{4}e^{-4\varphi}\mathbb F^T_{\mu\nu}\mathcal H\mathbb F^{\mu\nu},
\end{align}
where $\mathbb F_{(2)}=\dd\mathbb A_{(1)}$ and $\mathcal H^{-1}=\eta\mathcal H\eta$. The action then has a manifest $O(d+p,d)$ symmetry
\begin{equation}
    \mathcal H\to\qty(\Omega^{-1})^T\mathcal H\Omega^{-1},\qquad\mathbb A_{(1)}\to\Omega\mathbb A_{(1)},\qquad \Omega^T\eta\Omega=\eta.
\end{equation}
However, we may further dualize the field strengths
\begin{equation}
    \mathbb F_{\mu\nu}{}^M=e^{2\varphi}\epsilon_{\mu\nu\rho}\partial^\rho\xi_N\mathcal H^{NM}.
\end{equation}
The scalar fields can then be organized into a larger matrix
\begin{align}
    \mathcal M_{\mathcal{MN}}=\begin{pmatrix}
        \mathcal H_{MN}+e^{4\varphi}\xi_M\xi_N&\quad e^{4\varphi}\xi_M&\quad-\mathcal H_{MP}\xi^P-\frac{1}{2}e^{4\varphi}\xi_M\xi_P\xi^P\\
        e^{4\varphi}\xi_N&\quad e^{4\varphi}&\quad -\frac{1}{2}e^{4\varphi}\xi_P\xi^P\\
        -\mathcal H_{NP}\xi^P-\frac{1}{2}e^{4\varphi}\xi_N\xi_P\xi^P&\quad-\frac{1}{2}e^{4\varphi}\xi_P\xi^P&\quad e^{-4\varphi}+\mathcal \xi_P \mathcal H^{PQ}\xi_Q+\frac{1}{4}e^{4\varphi}\qty(\xi_P\xi^P)^2
    \end{pmatrix},
\end{align}
where we have defined an enlarged index $\mathcal M=\{M,+,-\}$. We also define the ${O(d+p+1,d+1)}$-invariant bilinear form
\begin{equation}
    \tilde\eta_{\mathcal{MN}}=\begin{pmatrix}
        \eta_{MN}&\quad 0&\quad 0\\0&\quad 0& \quad 1\\0&\quad 1&\quad 0
    \end{pmatrix}.
\end{equation}
Moreover, note that on-shell the three-form $h_{(3)}$ is determined by a constant, which we will set to zero (see~\cite{Eloy:2022vsq} for the generalization to a massive deformation). The action may then be rewritten as
\begin{equation}
    e^{-1}\mathcal L_3=R+\frac{1}{8}\Tr\qty(\partial_\mu\mathcal M\partial^\mu\mathcal M^{-1}),
\end{equation}
where $\mathcal M^{-1}=\tilde\eta\mathcal M\tilde\eta$. Thus, we see that the action has a manifest invariance under the $O(d+p+1,d+1)$ transformation
\begin{equation}
    \mathcal M\to\qty(\tilde\Omega^{-1})^T\mathcal M\tilde\Omega^{-1},\qquad\tilde\Omega^T\tilde\eta\tilde\Omega=\tilde\eta.
\end{equation}

Given the $O(d+p+1,d+1)$ generators $T_{\mathcal{MN}}$, we may decompose these into $O(d+p,d)$  components as
\begin{equation}
    T_{\mathcal{MN}}=\{T_{MN},T_{M+},T_{M-},T_{+-}\}.
\end{equation}
The $T_{MN}$ act as $O(d+p,d)$ T-duality transformations,
\begin{equation}
    \mathcal H_{MN}\to\Omega_M{}^P\Omega_N{}^Q\mathcal H_{PQ},\qquad\xi_M\to\Omega_M{}^N\xi_N,
\end{equation}
the $T_{M+}$ act as constant shifts of scalars, 
\begin{equation}
    \xi_M\to\xi_M+c_M,\label{eq:shiftSym}
\end{equation}
the $T_{M-}$ act as nontrivial solution-generating U-duality transformations, and $T_{+-}$ acts as an $O(1,1)$ scaling 
\begin{equation}
    e^\varphi\to\Lambda e^\varphi,\qquad \xi_M\to \Lambda^{-2}\xi_M.\label{eq:O88scale}
\end{equation}
In particular, the $O(d+p,d)$ symmetry originates from T-duality and will be preserved by higher-derivative corrections~\cite{Sen:1991zi,Hohm:2014sxa}, while the shifts~\eqref{eq:shiftSym} are (large) gauge symmetries and will hence also be preserved. However, the $O(1,1)$ scaling transformation will be broken by the higher-derivative corrections. This is easy to see as the leading four-derivative action in $d+3$ dimensions added to~\eqref{eq:het0} is given by\footnote{We are being agnostic as to whether we are talking about heterotic or bosonic supergravity, but, either way, there will be a Riemann-squared term.}
\begin{equation}
    \hat e^{-1}\mathcal L_{4\partial}=e^{-2\phi}\hat R_{\hat\mu\hat\nu\hat\rho\hat\sigma}\hat R^{\hat\mu\hat\nu\hat\rho\hat\sigma}+\cdots.
\end{equation}
This will reduce to a term of the form $e^{-4\varphi}R_{\mu\nu\rho\sigma}^2+\cdots$, which is not invariant under the $O(1,1)$ transformation~\eqref{eq:O88scale}. A likewise argument applies to the tree-level eight-derivative $e^{-2\phi}t_8t_8\hat R^4$ term present in the type II theory.

From the $O(d+p+1,d+1)$ commutation relations,
\begin{equation}
    [T_{\mathcal{MN}},T_{\mathcal{PQ}}]=\eta_\mathcal{MQ}T_\mathcal{NP}-\eta_\mathcal{NQ}T_\mathcal{MP}-\eta_\mathcal{MP}T_\mathcal{NQ}+\eta_\mathcal{NP}T_\mathcal{MQ},
\end{equation}
one finds that the $T_{MN}$ form an $O(d+p,d)$ subalgebra
\begin{equation}
    [T_{MN},T_{PQ}]=\eta_{MQ}T_{NP}-\eta_{NQ}T_{MP}-\eta_{MP}T_{NQ}+\eta_{NP}T_{MQ},
\end{equation}
while $T_{\pm M}$ transforms as $O(d+p,d)$ vectors
\begin{equation}
    [T_{MN},T_{\pm P}]=-\eta_{PM}T_{\pm N}+\eta_{PN}T_{\pm M},
\end{equation}
and $T_{+-}$ transforms as an $O(d+p,d)$ scalar
\begin{equation}
    [T_{MN},T_{+-}]=0.
\end{equation}
We also have the relation
\begin{equation}
    [T_{+M},T_{-N}]=\eta_{MN}T_{+-}-T_{MN}.
\end{equation}
Consequently, this last relation implies that all the $T_{-M}$ must be broken. In particular, this means that the global symmetry algebra is spanned by $T_{MN}$ and $T_{+M}$, which is equivalent to $O(d+p,d)\ltimes\mathbb R^{2d+p}$.

\section{Discussion}\label{sec:disc}
In this paper, we have shown that the breaking of the $\mathbb R^+$ scaling symmetry by higher-derivative corrections in minimal five-dimensional supergravity explicitly breaks all hidden symmetry enhancement of $G_{2(2)}$. Here, we have focused on the case of reducing along one spacelike and one timelike direction; however, an identical analysis applies to the case with two spacelike directions. We have also rederived that the higher-derivative corrections prevent all enhancement of the T-duality symmetry group $O(d+p,d)$ to $O(d+p+1,d+1)$ in heterotic and bosonic supergravity reduced to three dimensions on $T^d$. Note that heterotic supergravity in six dimensions, reduced on a circle, yields the STU model. As such, there is a special case of $d=3$ and $p=0$ in the $O(d+p+1,d+1)$ symmetry enhancement that corresponds to the $O(4,4)$ found in~\cite{Chong:2004na,Galtsov:2008bmt,Galtsov:2008jjb}. As such, our results imply that this symmetry enhancement is broken down to $O(3,3)\ltimes\mathbb R^{6}$. It is already well established that higher-derivative corrections should break \emph{some} of the U-duality symmetry; however, our key finding is that they break \emph{all} U-duality symmetry beyond T-duality, thus complicating application to higher-derivative solution generation.

It was already observed in~\cite{Eloy:2022vsq} that the higher-derivative corrections to heterotic and bosonic string theory prevent the symmetry enhancement from $O(d,d)$ to $O(d+1,d+1)$ in three dimensions.  However, the argument presented was based on a brute force dimensional reduction approach, where the preserved symmetry group was determined as the invariance group of an unphysical tensor compensator, whereas here we have made a group theoretic argument based on the structure of the underlying symmetry algebra. It should be emphasized that the result for $O(d+1,d+1)$ does not immediately imply the one for $G_{2(2)}$, as, although heterotic supergravity on $T^5$ can be truncated to five-dimensional minimal supergravity at the two-derivative level, this is no longer a consistent truncation at the four-derivative level~\cite{Liu:2023fqq}. Nevertheless, the same algebraic structure underlies the breaking of U-duality, although, strictly speaking, different scaling symmetries are broken in the two cases. We also point out that our argument includes the heterotic gauge fields in the $O(24,8)$ case, whereas Ref.~\cite{Eloy:2022vsq} truncates them, although these may always be reinstated via a consistent truncation~\cite{Xia:2025lvn}.

Given a (super)gravity theory with continuous global symmetry $G(\mathbb R)$, the quantization of the theory is expected to break this to its arithmetic subgroup $G(\mathbb Z)$. In the case of $O(d+p+1,d+1)$, this discretization to $O(d+p+1,d+1;\mathbb Z)$ trivializes the scaling, as noted in Ref.~\cite{Eloy:2022vsq}. This is because the $O(1,1)$ scaling transformation
\begin{equation}
    \tilde\Omega=\begin{pmatrix}
        \mathbbm{1}&\quad0&\quad 0\\0&\quad\Lambda&\quad0\\0&\quad0&\quad\Lambda^{-1}
    \end{pmatrix}\in O(d+p+1,d+1),
\end{equation}
is restricted to having integer entries, which restricts $\Lambda=\pm 1$. This trivializes the scaling to $O(1,1;\mathbb Z)\cong\mathbb Z_2$, which is a symmetry of the four-derivative action. As such, our group theoretic argument does not directly apply in the discrete case, although Ref.~\cite{Eloy:2022vsq} showed that the discrete symmetry enhancement is also fully broken. Similarly, in the $G_{2(2)}$ case, the $\mathbb R^+$ scaling~\eqref{eq:R+combo} acts, in the representation of~\cite{Compere:2009zh}, via ${\mathrm{diag}(\Lambda^{-2},\Lambda^{-1},\Lambda^{-1},1,\Lambda,\Lambda,\Lambda^2)\in G_{2(2)}}$ with $\Lambda\in\mathbb R^+$. Restricting entries to integers thus forces $\Lambda=1$, thereby trivializing the scaling. Nevertheless, we can make a slightly different argument. Note that the leading term in the action will reduce as
\begin{equation}
    \sqrt{-\hat g}\hat R_{\hat\mu\hat\nu\hat\rho\hat\sigma}R^{\hat\mu\hat\nu\hat\rho\hat\sigma}\to\sqrt{g}\tau R_{\mu\nu\rho\sigma}R^{\mu\nu\rho\sigma}+\cdots.
\end{equation}
Clearly, $\sqrt{g}R_{\mu\nu\rho\sigma}^2$ will be invariant under $G_{2(2)}$, which transforms only the moduli $\chi_i$ and $\vec\phi$.%
\footnote{In three dimensions, the Riemann tensor may be expressed in terms of the Schouten tensor, $S_{\mu\nu}=R_{\mu\nu}-\tfrac{1}{4}R\,g_{\mu\nu}$. This may be field redefined to replace $R_{\mu\nu}$ with $\Tr\qty(\partial_\mu\mathcal M\,\partial_\nu\mathcal M^{-1})$. Regardless, the resulting expression will still be $G_{2(2)}$ invariant.}
As such, the only transformations that will leave this action invariant are the ones that leave $\tau$ invariant. The $SL(2,\mathbb Z)$ large diffeomorphisms of the torus and the axion shifts $e_i$ will leave $\tau$ invariant, but it can be checked that all the $k_i$ ($i\ne 1$) do not. As such, we will still break $G_{2(2)}(\mathbb Z)$ to its geometric subgroup $SL(2,\mathbb Z)\ltimes\mathfrak{g}_{5,4}(\mathbb Z)$. A near-identical argument can be made to see that $O(d+p+1,d+1;\mathbb Z)$ must be broken to its ``geometric'' subgroup $O(d+p,d;\mathbb Z)\ltimes\mathbb Z^{2d+p}$.

Our results imply that U-duality fails to be present order-by-order in the higher-derivative corrections to non-maximal supergravity when reduced on a torus. Minimal supergravity in five dimensions can be constructed as a consistent truncation of eleven-dimensional supergravity on $T^6$. As such $G_{2(2)}$ will correspond to an \.In\"on\"u-Wigner contraction of $E_{8(8)}$. Seeing as any higher-derivative corrections fully break the symmetry enhancement to $G_{2(2)}$, this suggests that $E_{8(8)}$ must also be (at least partially) broken by the eight-derivative corrections. In particular, we expect that at least six of the generators of $E_{8(8)}$ (corresponding to $\vec\alpha_4\cdot\vec h$, $f_2$,..., $f_6$ of $G_{2(2)}$) should be broken. However, the $O(7,7)$ symmetry arising from T-duality persists to all orders in the tree-level $\alpha'$-expansion, and this contains a non-geometric $\beta$ symmetry~\cite{Baron:2022but,Baron:2023qkx}. As such, we do not expect the two-derivative $E_{8(8)}$ to be fully broken to the geometric subgroup of $SL(8,\mathbb R)\ltimes\mathbb R^{56}$ by tree-level higher-derivative corrections in IIA supergravity; however, it is not clear if any hidden symmetries beyond $O(7,7)$ (and shifts of the axions) may survive. Notably, Ref.~\cite{Lambert:2006he} found that the roots of $E_{8(8)}$ do not appear in the $T^8$ reduction of higher-derivative eleven-dimensional supergravity as one would expect, but rather a combination of roots and weights appears, which supports that the eight-derivative corrections break the $E_{8(8)}$ symmetry. However, tree-level $\mathcal R^4$ corrections in eleven-dimensional supergravity on a circle include one-loop $\mathcal R^4$ corrections in IIA supergravity, which will come with additional subtleties. It should be emphasized that U-duality is inherently a nonperturbative symmetry of the full quantum theory, so it is likely to be realized nontrivially. In particular, Ref.~\cite{Lambert:2006ny} suggested that the scalars should transform as non-holomorphic automorphic forms of $E_{8(8)}(\mathbb Z)$. By contraction, one might then expect that the $G_{2(2)}$ symmetry is realized if our scalars transform as automorphic forms of $G_{2(2)}(\mathbb Z)$. Morally, this is a replacement of our basic linear representations with an ``automorphic representation.''

Ultimately, the problem boils down to neglecting non-perturbative effects. In string theory, instantons generally take the form of a Euclidean brane wrapping a compact spacetime cycle~\cite{Robles-Llana:2006hby,Alexandrov:2008gh,Alexandrov:2009zh}. For example, the $SL(2,\mathbb Z)$ invariant IIB action~\eqref{eq:SL2invAction} receives contributions from tree-level, one-loop, and D-instanton effects. Without the D-instanton contributions, S-duality would fail to be preserved. On the other hand, there are no D$(-1)$-branes in the heterotic and bosonic theories, so there are no such instanton contributions in ten dimensions. Nevertheless, in dimension $D\le 4$, there will be instanton contributions from Euclidean NS5 branes wrapping six-cycles of the torus~\cite{Mohaupt:2000mj}. Of course, for the Type II case, in addition to NS5-instantons, there will also be instanton contributions from Euclidean D$p$-branes wrapping $(p+1)$-cycles of the torus. Such background dependence means that one effectively must compute the effective action for each particular background. Thus, U-duality (notwithstanding T-duality) cannot be efficiently used for generating classical higher-derivative black hole solutions in the original $(d+3)$-dimensional parent theory, unlike the case at the two-derivative level.

Such instanton contributions are not well understood, aside from the D$(-1)$ case. Nevertheless, for the case of IIB string theory reduced on $T^d$, the supersymmetric Ward identities and U-duality together are sufficient to fix the scalar factor in front of the $\mathcal R^4$ term to be a particular (regularized) Langlands-Eisenstein series associated to the maximal parabolic subgroup of $E_{d+1(d+1)}(\mathbb Z)$~\cite{Green:1997di,Kiritsis:1997em,Pioline:1997pu,Obers:1999um,Pioline:2010kb,Green:2010wi,Green:2011vz,Bossard:2015oxa,Bossard:2015foa} (See~\cite{Ozkan:2024euj} for a recent review.). This is expected to be equivalent to the contributions from instantons. For the non-maximal case, we conjecture that we should expect the appearance of an automorphic form of $G_{2(2)}(\mathbb Z)$ or $O(d+p+1,d+1;\mathbb Z)$ to appear in front of the action in three dimensions, which should incorporate the effects of instanton contributions. Such automorphic forms of $G_{2(2)}(\mathbb Z)$ are not currently well understood~\cite{pollack:2018}. Moreover, it is also not clear if there is enough supersymmetry in the non-maximal case to fix the automorphic form uniquely. Nevertheless, for both heterotic and minimal five-dimensional supergravity, supersymmetry uniquely fixes the tensorial structure of the action, which suggests that it may also be sufficient to fix the automorphic function that arises from instanton contributions.

Similarly, background dependence is expected to arise in $O(d+p,d)$ solution generation at the loop level. Classically, the coupling constants present in the higher-derivative string effective action are the same for any background. However, quantum corrections are inherently background dependent. When there is a spacetime torus, string loop S-matrix elements involve both KK momentum and winding modes, which correspondingly affect the higher-derivative couplings in the effective action~\cite{Green:1982sw}.\footnote{We remark that the tree-level corrections do not contain internal momenta, which is why the two-derivative action is not affected.} The presence of winding modes leads to additional contributions not present in non-compact spacetimes. As such, the quantum effective action for strings in a background with circular isometries cannot be obtained from the dimensional reduction of the Minkowski effective action. This is true even at the level of T-duality: The one-loop effective action reduced on a circle will not have an $O(1,1)$ symmetry~\cite{Garousi:2013qka}. This is to say that the ten-dimensional and lower-dimensional effective actions are no longer equivalent at the quantum level. However, this will not affect the four-derivative corrections as, even incorporating winding mode contributions, the heterotic one-, two-, and three-point functions vanish at the loop level, for all genera, due to kinematic constraints~\cite{Green:1982sw}. This is to say that there are great difficulties with extending solution generation to the one-loop action. Fortunately, while the contribution of winding modes may interfere at higher-derivative orders, it will not affect the four- and six-derivative heterotic effective action.

\let\oldaddcontentsline\addcontentsline% Store \addcontentsline
\renewcommand{\addcontentsline}[3]{}% Make \addcontentsline a no-op
\begin{acknowledgments}
    This work is supported by the National Natural Science Foundation of China (NSFC) under Grants No. 12175164 and No. 12247103.
\end{acknowledgments}
\let\addcontentsline\oldaddcontentsline% Restore \addcontentsline

\let\oldaddcontentsline\addcontentsline% Store \addcontentsline
\renewcommand{\addcontentsline}[3]{}% Make \addcontentsline a no-op
\bibliography{cite}
\let\addcontentsline\oldaddcontentsline% Restore \addcontentsline

\end{document}